\documentclass[12pt,preprint]{aastex}
\usepackage{natbib}

\shorttitle{Low-metallicity host of SN~2010jl}
\shortauthors{Stoll et al.}

\begin{document}

\title{SN~2010jl in UGC~5189:  Yet Another Luminous Type IIn Supernova in a Metal-Poor Galaxy}

\author{R.~Stoll\altaffilmark{1}, J.L.~Prieto\altaffilmark{2,3}, 
        K.Z.~Stanek\altaffilmark{1,4}, R.W.~Pogge\altaffilmark{1,4}, 
         D.M.~Szczygie{\l}\altaffilmark{1}, G.~Pojma{\'n}ski\altaffilmark{5},
        J.~Antognini\altaffilmark{1}, H.~Yan\altaffilmark{4}}

\email{stoll@astronomy.ohio-state.edu}

\altaffiltext{1}{Department of Astronomy, The Ohio State University, 
                 140 W. 18th Ave., Columbus OH 43210}
\altaffiltext{2}{Carnegie Observatories, 813 Santa Barbara Street, 
                 Pasadena, CA 91101}
\altaffiltext{3}{Hubble and Carnegie-Princeton Fellow}
\altaffiltext{4}{Center for Cosmology and AstroParticle Physics, 
                 The Ohio State University, 191 West Woodruff Avenue, 
                 Columbus OH 43210}
\altaffiltext{5}{Warsaw University Astronomical Observatory, 
                 Al.~Ujazdowskie 4, 00-478 Warsaw, Poland}

\begin{abstract}
We present All-Sky Automated Survey data starting 25 days before the 
discovery of the recent type 
IIn SN~2010jl, and we compare its light curve to other luminous IIn 
SNe, 
showing that it is a luminous ($M_I \approx -20.5$) event. 
Its host galaxy, UGC~5189, has a low gas-phase 
oxygen abundance ($\rm 12 + log(O/H) = 8.2\pm0.1$), which reinforces the 
emerging trend that over-luminous core-collapse supernovae are 
found in the low-metallicity tail of the galaxy distribution, similar to 
the known trend for the hosts of long GRBs.  We compile 
oxygen abundances from the literature and from our own observations of 
UGC~5189, and we present an unpublished spectrum of the luminous type Ic 
SN~2010gx that we use to estimate its host metallicity.  
We discuss these in the context of host metallicity trends for different
classes of core-collapse objects.  
The earliest generations of stars are known to be enhanced in [O/Fe] relative 
to the Solar mixture; it is therefore likely that the stellar progenitors of 
these 
overluminous supernovae are even more iron-poor than they are oxygen-poor.
A number of mechanisms and massive star progenitor systems 
have been proposed to explain the most luminous core-collapse 
supernovae. Any successful theory that tries to explain these very luminous 
events will need to include the emerging trend that points towards 
low-metallicity for the massive progenitor stars.
This trend for very luminous 
supernovae to strongly prefer low-metallicity galaxies should be taken into 
account when considering various aspects of the evolution of the 
metal-poor early universe, 
such as enrichment and reionization.

\end{abstract}


\section{Introduction}\label{sec:intro}


The bright SN~2010jl, discovered on UTC 2010 November 3.5 \citep{newton10} 
was classified as a type IIn supernova \citep{benetti10,yamanaka10}.  
A possible luminous blue progenitor has been 
identified  in archival HST WFPC2 data \citep{smith10jl}.  
If this detection is of a massive 
compact cluster, the turnoff mass is constrained to be $> 30$~M$_\odot$.  
If it is a single star, it is either a massive $\eta$~Carinae-like star or 
something fainter that has been caught in an LBV-like eruption phase, possibly 
a precursor explosion.
Spectropolarimetric 
observations indicate possible asymmetry in the explosion geometry and 
limit the amount of dust in the progenitor environment 
\citep{patat10}.

\citet{prieto10} point out that the supernova's host galaxy, UGC~5189, has 
been shown 
to be metal-poor based on a spectrum obtained by the SDSS survey a few 
arcseconds away from the site of the supernova.  It is included 
in the \citet{tremonti04} DR4 catalog of galaxy 
metallicities, which estimates 
$\rm 12 + log(O/H) = 8.15\pm 0.1$~dex based on strong recombination and 
forbidden 
emission lines in the spectrum.  \citet{pilyugin07} estimate 
$\rm 12 + log(O/H) = 8.3$~dex from the same spectrum based on the ``direct'' 
photoionization-based abundance method using an estimate of the electron 
temperature ($T_e$) from the [\ion{O}{3}]$\lambda 4363$\AA{} auroral line.


Throughout this paper, as above, we will adopt O/H as a proxy for the 
``metallicity'' of the host galaxy, because only gas-phase oxygen abundance 
estimates are available.  In Section~\ref{sec:discuss} we will discuss how 
this relates to total metal abundance and iron abundances.  

There is mounting evidence showing that the majority of the most 
optically luminous 
supernovae explode in low-luminosity, star-forming galaxies, which are 
likely to be metal-poor environments.  
Circumstantial 
evidence comes from the fact that most of these objects have been discovered 
in new rolling searches that are not targeted to bright galaxies, even 
though they were bright enough to be discovered in galaxy-targeted searches
(the only exception is the recently discovered SN~2010jl).  Some of the 
rolling searches that are discovering the most energetic supernovae include 
the Texas Supernova Search \citep[TSS;][]{quimbyTSS}, the Catalina Real-Time 
Transient Survey \citep[CRTS;][]{drake09}, 
the Palomar Transient Factory \citep[PTF;][]{rau09}, and the Panoramic 
Survey Telescope \& Rapid Response System (Pan-STARRS).
\citet{neill10} find that luminous SNe occur predominantly in the faintest, 
bluest galaxies.  \citet{li10} find that SNe IIn may preferentially occur in 
smaller, less-luminous, later-type galaxies than SNe II-P.  
The mass-metallicity relationship implies that such small, faint 
galaxies should tend to be low-metallicity \citep{tremonti04}.

\citet{kozlowski10} presented the first host galaxy luminosity vs. oxygen 
abundance diagram with a small sample of three energetic type IIn and type Ic 
core-collapse supernovae (CCSNe)
compiled from published work in the literature (two objects) and new data 
for SN~2007va.
These initial results suggest that the host galaxy environments of the 
most energetic CCSNe are on average metal-poor (metallicities 
$\sim 0.2 - 0.5$~$Z_\odot$) compared to the bulk of star-forming galaxies 
in SDSS, and are similar in metallicity to the hosts of local GRBs 
\citep{stanek06}.  It has now been 
well-established that long GRBs are accompanied by broad-line type Ic SNe 
\citep[e.g.][]{stanek03}, firmly connecting these very energetic explosions 
to the deaths of massive stars.  
In contrast, luminous type IIn SNe are not shown to be connected with local 
GRBs \citep[although see][]{germany00,rigon03}.
The sample of luminous supernovae with 
measured abundances is small and incomplete because only some of 
the brightest host galaxies have been targeted, which hinders the 
interpretation and comparison of the results.

In this paper we expand this sample with metallicities of the hosts of 
SN~2010gx and SN~2010jl and put all metallicity measurements on a common 
scale to confirm the emerging trend that these optically luminous 
core-collapse events appear to occur in low-metallicity or low-luminosity 
hosts.

\section{Observations}

\subsection{Photometry}\label{sec:photobs}
Photometric data were obtained with the 10-cm All-Sky 
Automated Survey (ASAS) North 
telescope in Hawaii \citep{pojmanski02,pigulski09}.  These data were 
processed with the reduction pipeline described 
in detail in \citet{pojmanski98,pojmanski02}.  The $V$ magnitudes are 
tied to the Johnson $V$ scale using Tycho \citep{hog00} and 
Landolt \citep{landolt83} standard stars.  
The $I$ magnitudes were calibrated using transformed SDSS DR7 \citep{dr7}
magnitudes of standards in the field.  The magnitudes were measured using 
aperture photometry with a 2 pixel (30 arcsec) aperture radius.  We subtract 
the contribution from the host galaxy measuring the median magnitude of the 
host in the same aperture using all archival ASAS images before the 
supernova, which give $V_{host} = 14.97$~mag and $I_{host} = 14.28$~mag.

We obtained $UBVRI$ images of SN~2010jl with the SITe-3 CCD camera 
mounted on the Swope 1-m telescope at Las~Campanas Observatory on three 
consecutive nights from UT November $15-17$ 2010.  The images were reduced 
with standard tasks in the IRAF\footnote{IRAF 
is distributed by the National Optical Astronomy Observatory, which is 
operated by the Association of Universities for Research in Astronomy (AURA) 
under cooperative agreement with the National Science Foundation.} 
ccdproc package.  The data were flat-fielded using domeflats obtained for 
$BVRI$ images and twilight skyflats for $U$-band images.  We applied a 
linearity correction using the prescription and coefficients from 
\citet{hamuy06}.  The photometry of the supernova was obtained using 
aperture photometry and a $2\arcsec$ radius aperture with the phot task 
in IRAF.  The magnitudes were calibrated using field stars with 
SDSS DR7 $ugriz$ photometry, converted to standard $UBVRI$ magnitudes 
through standard equations on the SDSS website.  We subtracted the 
contribution from the host galaxy fluxes measuring the magnitudes in the 
same aperture using the pre-explosion SDSS images.  
In Table~\ref{table:phot} we gather all optical 
photometry presented in this paper.

\begin{deluxetable}{lcccccccccccc}
\tablecolumns{12}
\tabletypesize{\scriptsize}
\tablewidth{0pt}
\tablecaption{Optical photometry of SN~2010jl\label{table:phot}}
\tablehead{
   \colhead{HJD}                    &
   \colhead{U}                      &
   \colhead{$\sigma$}               &
   \colhead{B}                      &
   \colhead{$\sigma$}               &
   \colhead{V}                      &
   \colhead{$\sigma$}               &
   \colhead{R}                      &
   \colhead{$\sigma$}               &
   \colhead{I}                      &
   \colhead{$\sigma$}               &
   \colhead{Telescope /}            \\
   \colhead{-2450000}               &
   \colhead{(mag)}                  &
   \colhead{}                       &
   \colhead{(mag)}                  &
   \colhead{}                       &
   \colhead{(mag)}                  &
   \colhead{}                       &
   \colhead{(mag)}                  &
   \colhead{}                       &
   \colhead{(mag)}                  &
   \colhead{}                       &
   \colhead{Instrument}                       
}
\startdata
5479.14 & \nodata & \nodata & \nodata & \nodata & 13.79 & 0.10 & \nodata & \nodata & \nodata & \nodata & ASAS \\ 
5484.13 & \nodata & \nodata & \nodata & \nodata & 13.73 & 0.10 & \nodata & \nodata & 13.28 & 0.10 & ASAS \\ 
5488.12 & \nodata & \nodata & \nodata & \nodata & 13.67 & 0.10 & \nodata & \nodata & 13.09 & 0.10 & ASAS \\ 
5493.14 & \nodata & \nodata & \nodata & \nodata & 13.73 & 0.10 & \nodata & \nodata & \nodata & \nodata & ASAS \\ 
5494.12 & \nodata & \nodata & \nodata & \nodata & \nodata & \nodata & \nodata & \nodata & 13.00 & 0.10 & ASAS \\ 
5504.10 & \nodata & \nodata & \nodata & \nodata & 13.78 & 0.10 & \nodata & \nodata & 13.10 & 0.10 & ASAS \\ 
5509.10 & \nodata & \nodata & \nodata & \nodata & 13.86 & 0.10 & \nodata & \nodata & \nodata & \nodata & ASAS \\ 
5509.13 & \nodata & \nodata & \nodata & \nodata & \nodata & \nodata & \nodata & \nodata & 13.18 & 0.10 & ASAS \\ 
5515.75 & 13.83 & 0.05 & 14.24 & 0.05 & 13.46 & 0.05 & 13.85 & 0.05 & 13.19 & 0.05 & Swope \\
5516.75 & 13.83 & 0.05 & 14.26 & 0.05 & 13.45 & 0.05 & 13.87 & 0.05 & \nodata & \nodata & Swope \\
5517.13 & \nodata & \nodata & \nodata & \nodata & \nodata & \nodata & \nodata & \nodata & 13.28 & 0.10 & ASAS \\ 
5517.75 & 13.86 & 0.05 & 14.26 & 0.05 & \nodata & \nodata & 13.87 & 0.05 & 13.20 & 0.05 & Swope \\
\enddata
\end{deluxetable}


\subsection{Spectroscopy}

\begin{figure}
  \begin{center}
  \begin{tabular}{c}
  \includegraphics[width=16cm]{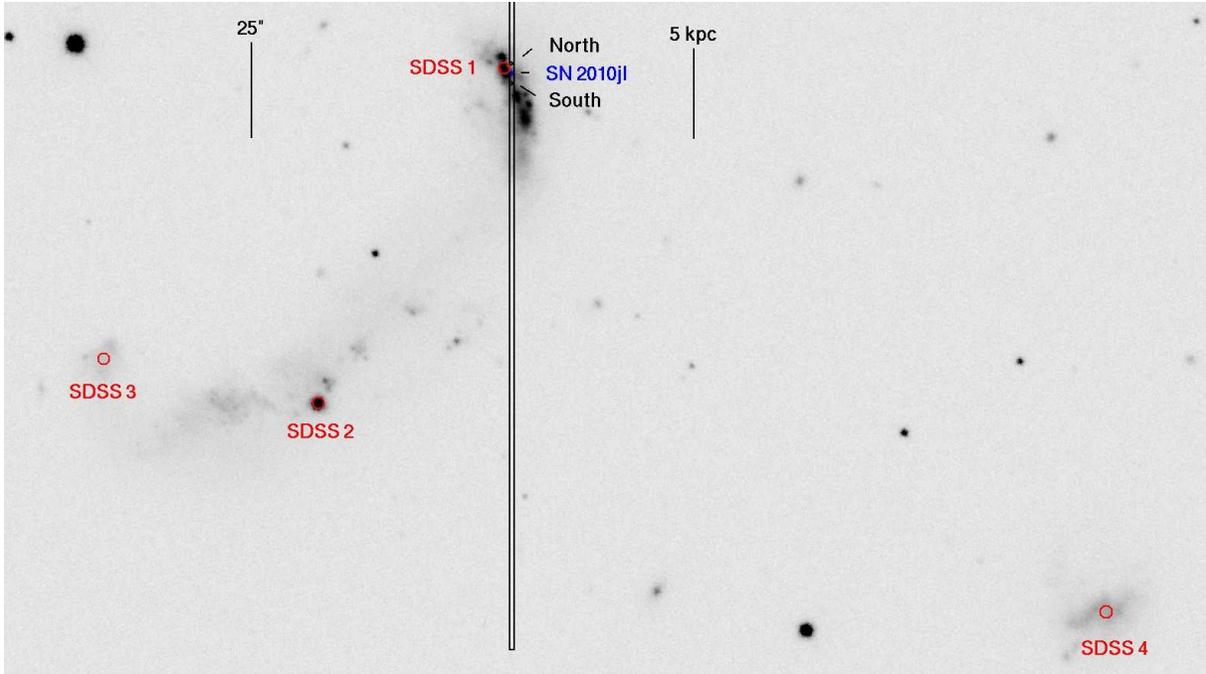}
  \end{tabular}
  \end{center}
  \caption 
   { \label{fig:finding}
      A schematic of the host galaxy of SN~2010jl.  North is up and east is 
      to the left.  The OSMOS slit position 
      is marked with a black rectangle 1.2 arcsec wide.  The slit was 
      centered on the supernova.  The fiber positions of archival SDSS 
      spectra of the galaxy are marked with the red circles, which have a 
      diameter of 3 arcseconds.  The background is constructed from SDSS 
      g imaging.  The small blue galaxy in the southwest corner of the image 
      has a similar spectrum and is at the same redshift (z=0.0107).  
      It is almost 
      certainly part of the interacting system.  The position of SN~2010jl 
      is marked, as well as the emission regions to the north and south  
      for which we extract OSMOS spectra.     
   }
  \end{figure}

Spectra were obtained on November 5 and November 11 2010 with the Ohio State 
Multi-Object Spectrograph \citep[OSMOS,][]{martini10,stoll10} on the 2.4-m 
Hiltner telescope.  The OSMOS slit was oriented N/S.  Although the instrument 
supports a 20 arcmin longslit, we observed the target with the CCD readout 
in a subframe that limited the effective slit length to 5~arcmin to reduce 
the readout time.  A schematic of the slit 
coverage on the field is shown in Figure~\ref{fig:finding}.  The 1.2 arcsec 
slits we used in the OSMOS observations are shown with black rectangles.  
The irregular galaxy has four 
archival SDSS spectra; the locations of the SDSS fiber apertures are shown 
with red circles.

We processed the raw CCD data using standard techniques in IRAF, including 
cosmic ray rejection using L.A.Cosmic \citep{vandokkum01}.  Each source was 
extracted individually with apall, extracting a sky spectrum simultaneously.  
The range of extraction apertures was $14-24$~pixels, corresponding to 
$3.8-6.6$~arcsec, or $0.8-1.4$~kpc at the distance of UGC~5189.
With the 1.2 arcsec slit used in our observations, the OSMOS VPH grism 
provides $R \sim 1600$ at 5000 \AA.  The red slit position we used 
provides a wavelength coverage of $3960-6880$~\AA.  
Wavelength calibration was done with Ar arclamp observations, and a 
spectrophotometric standard from \citet{oke90} was observed each night, with 
which we performed relative flux calibrations.

\begin{figure}
  \begin{center}
  \begin{tabular}{c}
  \includegraphics[width=16cm]{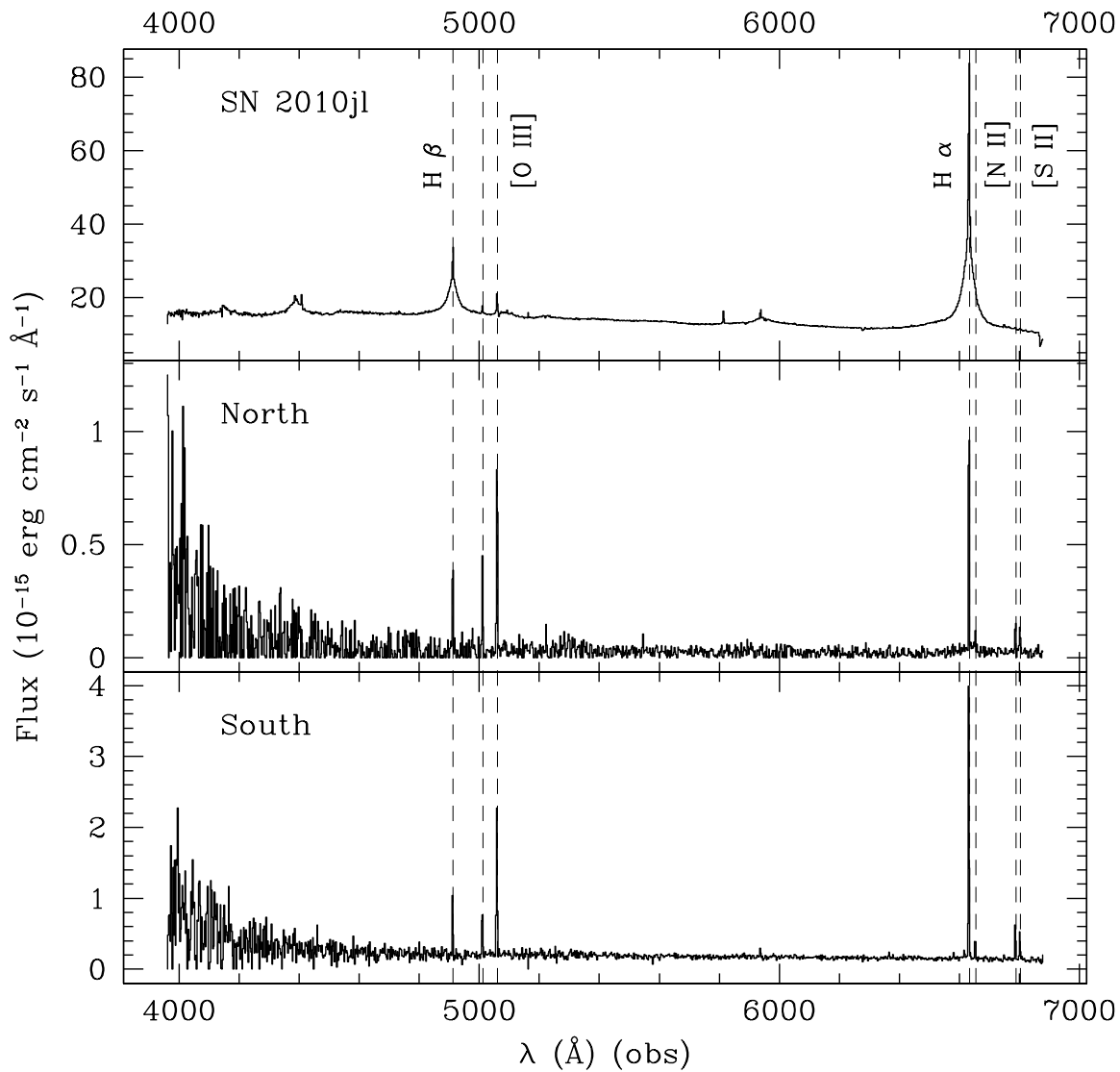}
  \end{tabular}
  \end{center}
  \caption 
   { \label{fig:redspec}
     OSMOS spectra of SN2010jl and the emission line regions $\sim$7 arcsec 
     ($\sim$1.4 kpc) 
     to the north and south.  Negative flux values from noise have been 
     truncated to zero.  Dashed lines show the 
     wavelengths of  H$\beta$~$\lambda 4861$,  
     [\ion{O}{3}]$\lambda 4959,5007$, H$\alpha$~$\lambda 6563$, 
     [\ion{N}{2}]$\lambda 6584$, and [\ion{S}{2}]$\lambda 6717,6731$
     at $z=0.0107$.
   }
  \end{figure}

The top panel of Figure~\ref{fig:redspec} shows a spectrum of SN~2010jl 
obtained 
on Nov. 5.  Taking advantage of the N/S slit, we extract host galaxy spectra 
centered $\sim 7$~arcsec on either side of the SN, corresponding to 
$\sim 1.5$~kpc at the distance of UGC~5189.  
In Section~\ref{strongline} we use strong-line abundance diagnostics on these 
host galaxy spectra to 
estimate the gas-phase oxygen abundances of the progenitor environment.

\begin{figure}
  \begin{center}
  \begin{tabular}{c}
  \includegraphics[width=13cm]{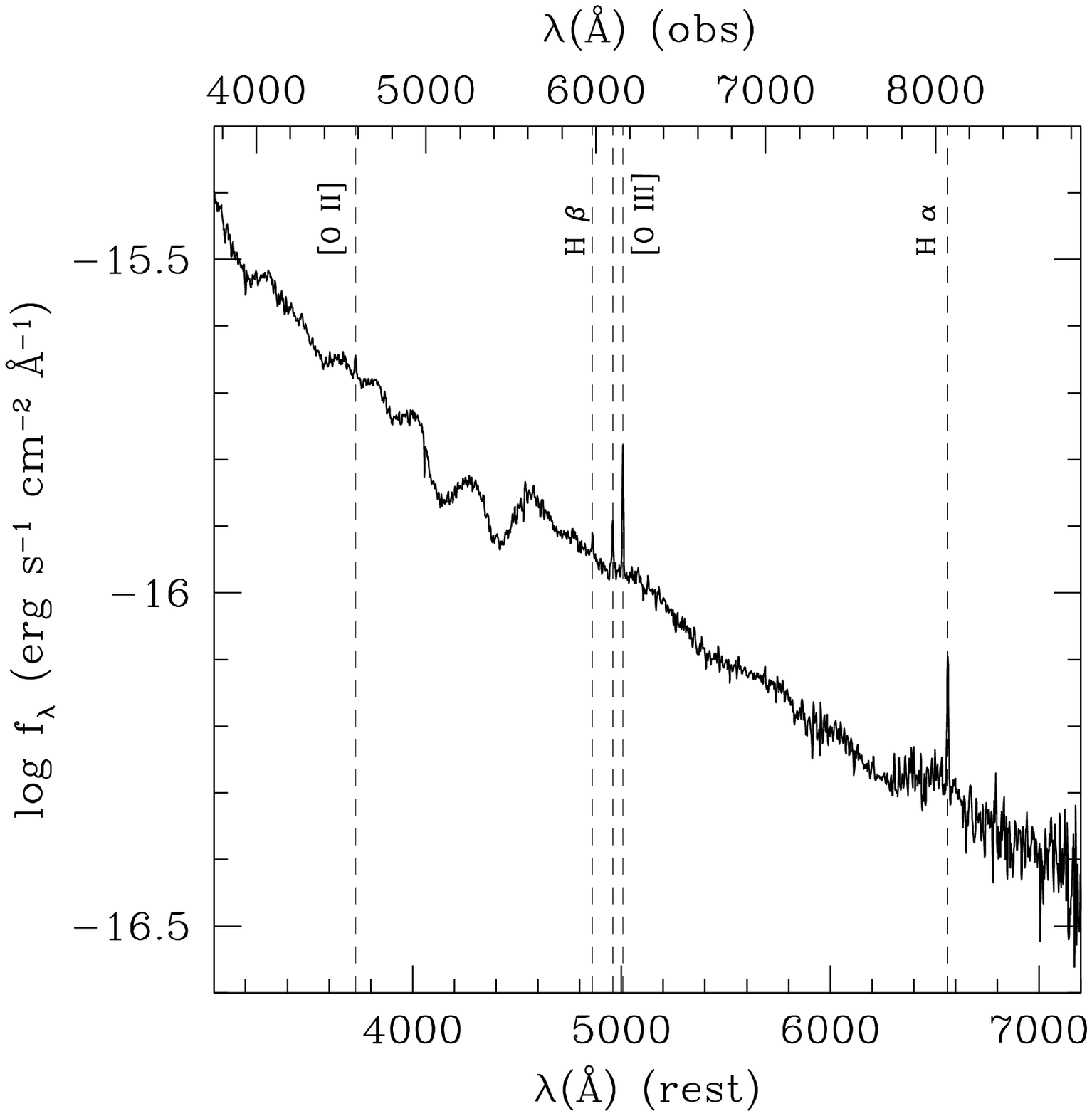}
  \end{tabular}
  \end{center}
  \caption 
   { \label{fig:gx}
    Spectrum of the luminous SN~2010gx obtained with the du Pont 2.5-m 
    telescope at the Las Campanas Observatory in March 2010, before the 
    supernova reached maximum light.  The spectrum has a blue continuum with 
    broad ($\sim 10000$~km~s$^{-1}$) absorption features likely associated 
    with ionized oxygen, nitrogen, and carbon.  Although the nature of these 
    events is still under debate, they share characteristics that are similar 
    to the spectra of SN~Ib/c \citep{quimby09,pastorello10}.  The dashed lines 
    mark the wavelengths of narrow emission lines of 
     [\ion{O}{2}]$\lambda 3727$, 
     H$\beta$~$\lambda 4861$,  
     [\ion{O}{3}]$\lambda 4959,5007$, and H$\alpha$~$\lambda 6563$ 
    detected from the star-forming host galaxy.  
    Using these lines, we can estimate an oxygen abundance for the host of 
    SN~2010gx of $\rm 12 + log(O/H) = 8.36$, using the strong line 
    diagnostic of \citet{kk04}.
    This corresponds to $\sim 0.3$~$Z_{\odot}$, similar to the 
    metallicity of the LMC.  In subsequent figures we convert this value to 
    the scale of the N2 diagnostic of \citet{pp04}, giving 
    $\rm 12 + log(O/H) = 8.16$.
   }
  \end{figure}

We also present in this paper a previously unpublished spectrum of 
the luminous ($M_B \approx -21.2$) type Ic SN~2010gx \citep{pastorello10} 
obtained with the WFCCD imager and 
spectrograph on the du~Pont 2.5-m telescope at Las~Campanas Observatory 
using a longslit mask (slit width $1.7\arcsec$) and the 300~l/mm grism.
We obtained $1200$-sec spectra on three consecutive nights 
(UT March $22-24$ 2010), or $4 \pm 2$~days (rest~frame) prior to the $B$-band 
peak.  
The WFCCD data were reduced using standard tools 
in IRAF.  We derived the wavelength solutions with HeNeAr arclamps 
obtained after each spectrum and did independent relative flux calibration 
using spectrophotometric standards observed each night.  
The final 
spectrum in Figure~\ref{fig:gx} is a combination of the three 
individual exposures, obtained in order to increase the S/N.  The 
spectrum covers the wavelength range $3700-9200$~\AA{} and has a FWHM 
resolution of $\sim 7$~\AA.  
Assuming a Galactic reddening law with $R_V = 3.1$, 
the total $V$-band extinction 
was estimated to be $A_V = 0.27$~mag using the ratio of H$\alpha$ to H$\beta$ 
fluxes and correcting to the theoretical case-B 
recombination value of 2.85.  The spectrum was corrected for this reddening 
before calculating line ratios.  The continuum was subtracted locally 
to each emission line before measuring the flux; the effect of the SN 
contamination on the metallicity determination should be small.
We estimate the oxygen abundance to be $\rm 12 + log(O/H) = 8.36$ using the 
diagnostic of \citet{kk04}, which uses 
$R_{23} =$~([\ion{O}{2}]$\lambda 3727 +$[\ion{O}{3}]$\lambda\lambda 
4959, 5007$)/H$\beta$ and 
$O_{32} =$~([\ion{O}{3}]$\lambda 5007 +$[\ion{O}{3}]$\lambda 4959)/$[\ion{O}{2}]$\lambda 3727$.  
These are highly sensitive to reddening, so the correction is crucial.
(In subsequent discussion we convert this metallicity to the scale of 
\citet{pp04} for uniformity.)

\section{Results and analysis}

\subsection{SN~2010jl light curve}

\begin{figure}
  \begin{center}
  \begin{tabular}{c}
  \includegraphics[width=14cm]{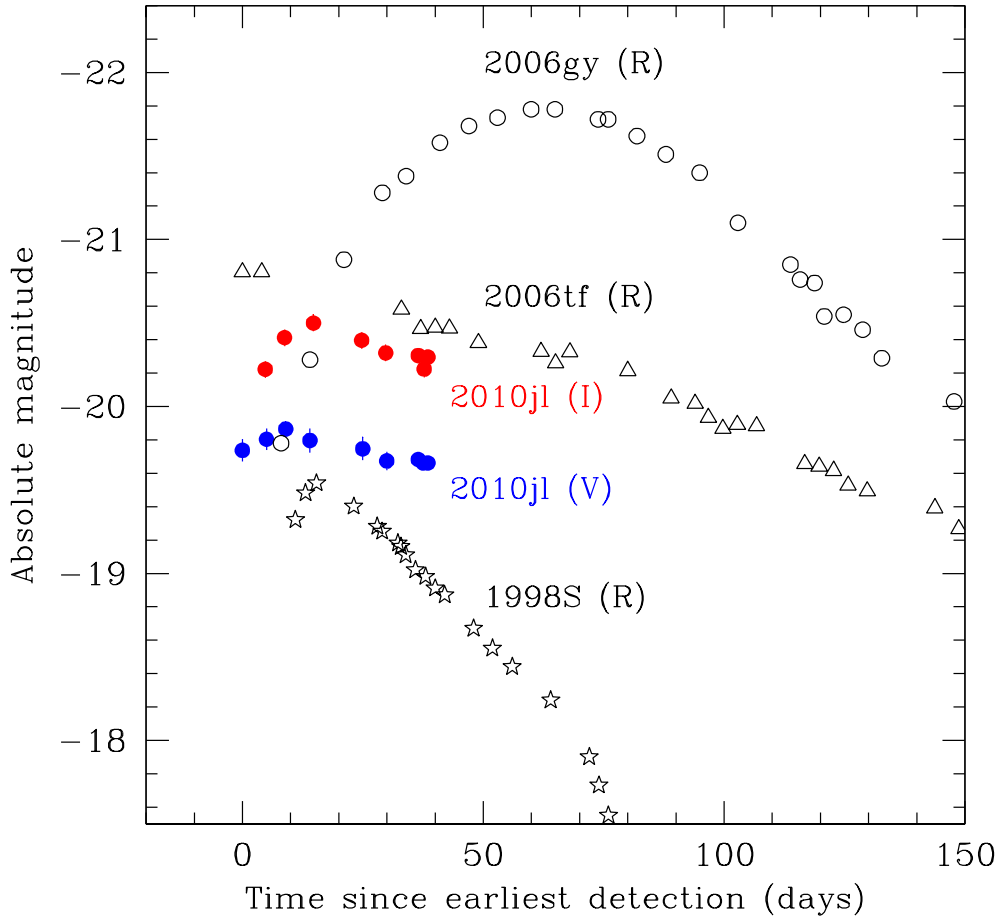}
  \end{tabular}
  \end{center}
  \caption 
   { \label{fig:lc}
      Absolute magnitude light curve of SN~2010jl, plotted for comparison 
      with other type IIn supernovae from the literature.  The filled circles 
      show the $V$ and $I$-band light curves of SN~2010jl from ASAS North 
      and Swope.   
      For all the other SNe we show the $R$-band light curves: 
      the ultraluminous SN~2006gy 
      \citep[open circles,][]{smith07gy} and 
      2006tf \citep[open triangles,][]{smith08tf}, and the type IIn 
      1998S \citep{fassia00}.  The normal range of peak luminosities for 
      type IIn SNe is $-18.5 \lesssim M_R \lesssim -17$ \citep{kiewe10}.  
      The horizontal axis is the time with 
      respect to the earliest detection.  In the case of SN~2010jl, the 
      SN was discovered on Nov 3.5, 2010 \citep{newton10}, but the first 
      detection from ASAS North is from Oct 9.6, 2010, 25 days earlier.
      }
  \end{figure}

The ASAS light curves of SN~2010jl include several pre-peak observations 
beginning 25 days before discovery.  We plot them in Figure~\ref{fig:lc} 
along with the $V$ and $I$-band Swope data.  For comparison, we plot 
$R$-band light curves from the literature of the ultraluminous 
SN~2006gy \citep{smith07gy} and SN~2006tf \citep{smith08tf}, and 
the luminous type IIn SN~1998S \citep{fassia00}.  
UGC~5189 was behind the Sun and unobservable prior to 
the first ASAS observations.  The color of the supernova changed from 
$V-I \approx 0.4$~mag at the epoch of ASAS discovery to $V-I \approx 0.7$~mag 
at the latest epochs.  The $UBVRI$ photometry from Swope is consistent 
with a $\sim 7000$~K blackbody, as \citet{smith10jl} conclude from a 
fit to the spectrum.
The normal range for peak luminosities of type IIn SNe is 
$-18.5 \lesssim M_R \lesssim -17$ \citep{kiewe10}.
The pre-discovery ASAS data constrain the 
peak luminosity of SN~2010jl ($M_V \approx -19.9$, $M_I \approx -20.5$), 
placing it firmly in the class of luminous (peak $M_R < -20$) type IIn 
supernovae.

\subsection{Metallicity near SN~2010jl}\label{strongline}

We use the empirical O3N2 and N2 diagnostics of \citet{pp04} and 
the theoretical and empirical N2 diagnostic of \citet{d02} to determine the 
metallicity of the OSMOS-observed \ion{H}{2} regions south and north of 
SN~2010jl.  
We choose these diagnostics based on the S/N and wavelength coverage of our 
observations.  The N2 diagnostic of \citet{pp04} and of \citet{d02} each 
depend solely on  
[\ion{N}{2}]$\lambda 6584$/H$\alpha \lambda 6563$.  This ratio is very 
insensitive to reddening due to the proximity of the lines.
The O3N2 diagnostic of \citet{pp04} also depends on 
[\ion{O}{3}]$\lambda 5007$/H$\beta \lambda 4861$, another closely-spaced 
pair of lines.  Although we do not correct these spectra for reddening, 
any reddening effect is very small compared to the scatter in each abundance 
diagnostic.
For subsequent analysis in Section~\ref{ztrends}, we convert metallicities of other galaxies 
determined using other diagnostics to these scales to enable direct 
comparison, using the empirical conversions of \citet{kewley08}.

For the galactic \ion{H}{2} regions 
$\sim 7$~arcsec ($\sim 1.5$~kpc) south and north of the SN site, 
using the \citet{pp04} O3N2 diagnostic, we find 
$\rm 12 + log(O/H) = 8.2 \pm 0.1$~dex.  Using the N2 diagnostic, we find 
$\rm 12 + log(O/H) = 8.2 \pm 0.1$~dex for the \ion{H}{2} region north of the 
SN 
and $8.3 \pm 0.1$~dex for the one to the south.  Using the 
\citet{d02} diagnostic, we find $\rm 12 + log(O/H) = 8.2 \pm 0.1$~dex and 
$8.4 \pm 0.1$~dex for the north and south regions, respectively.  

Our estimates of the oxygen abundance are consistent with previously 
published estimates for 
UGC~5189 based on the SDSS spectrum from $\sim 5$~arcsec ($\sim 1$~kpc) 
to the northeast of the SN site.  \citet{tremonti04} report 
$\rm 12 + log(O/H) = 8.15\pm 0.1$~dex for this galaxy, based on strong 
recombination and forbidden emission lines.  \citet{pilyugin07} report 
$\rm 12 + log(O/H) = 8.3$~dex from the same spectrum based on direct 
electron temperature using the [\ion{O}{3}]$\lambda 4363$~auroral line.  
For our analysis in Section~\ref{ztrends} we convert the \citet{tremonti04} 
metallicities, but for lack of an established empirical conversion, we omit 
the \citet{pilyugin07} metallicities.

\subsection{Metallicity elsewhere in the host galaxy}

There are two other SDSS spectra of the galaxy 
(see Figure~\ref{fig:finding}), including a 
bright point source in the south, approximately 20 kpc from the location of 
the supernova, marked in Figure~\ref{fig:finding} as SDSS~2.  
\citet{pilyugin07} report $\rm 12 + log(O/H) = 8.3$~dex for 
this spectrum.
The detection of \ion{He}{2}\,$\lambda 4686$\AA{} indicates that the ionizing 
continuum is hard \citep[e.g.][]{garnett91},  outside the regime where the 
\citet{pp04} and \citet{d02} diagnostics 
are valid.  \citet{brinchmann08} include this spectrum in their SDSS-based 
study of nearby galaxies with Wolf-Rayet features.  
\citet{tremonti04} do not fit a metallicity to this 
spectrum despite the high S/N, presumably because of 
the peculiar line ratios.

The other SDSS spectrum is of the eastern-most clump of UGC~5189, 
approximately 30 projected kpc from the location of the supernova 
(marked in Figure~\ref{fig:finding} as SDSS~3).  
Neither \citet{tremonti04} or \citet{pilyugin07} derive metallicities from 
this spectrum, which has lower S/N.  We find 
$\rm 12 + log(O/H) = 8.2\pm 0.1$~dex with all three line ratio 
diagnostics using the SDSS spectrum.  

Approximately 40 projected kpc from SN~2010jl is another blue galaxy at the 
same redshift (0.0106), which also has an SDSS spectrum, marked in 
Figure~\ref{fig:finding} as SDSS~4.  The redshift of the 
spectrum from the supernova site is 0.0107; these redshifts are consistent 
within the estimated 0.0001 redshift error for this lower S/N spectrum.  
From the shape and orientation of the 
wildly irregular UGC~5189, it seems very likely that this neighbor is a part 
of the interacting system.  Using the SDSS spectrum, we find 
$\rm 12 + log(O/H) = 8.2 \pm 0.1$~dex with both \citet{pp04} diagnostics and 
$8.3 \pm 0.1$~dex with \citet{d02}.  

UGC~5189 appears to be a high-mass, metal-poor outlier from the 
mass-metallicity relationship \citep[e.g.][]{peeples09}.  
Such similar metallicity measurements from widely spaced locations and 
widely varying diagnostics provide reassurance that there are no significant 
metallicity variations in UGC~5189, and that the three metallicity 
measurements described in Sec.~\ref{strongline} from a kpc or more away 
from the supernova site are likely to 
well-describe the supernova progenitor region.  All evidence points toward 
a SN~2010jl metallicity of $\lesssim 0.3$~Z$_\odot$.

\subsection{Host metallicity trends}\label{ztrends}

\begin{figure}
  \begin{center}
  \begin{tabular}{c}
  \includegraphics[width=13cm]{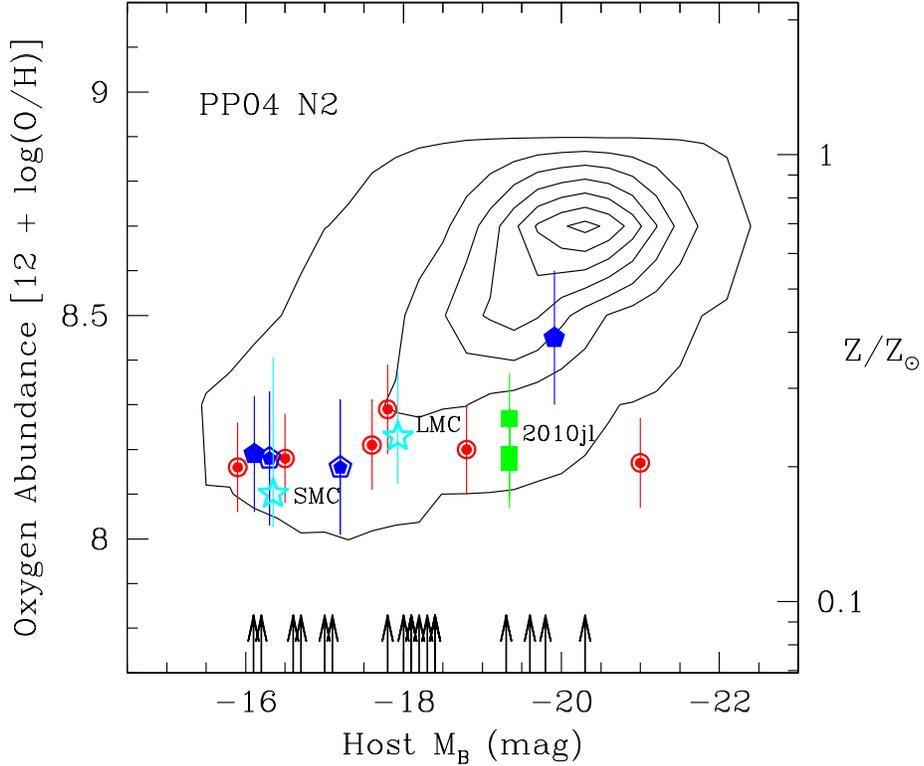}
  \end{tabular}
  \end{center}
  \caption 
   { \label{fig:ZBpp04n2}
      Host galaxy metallicities and absolute magnitudes are plotted for 
      luminous SNe (blue pentagons) including SN~2010jl (green squares).
      The luminous supernovae without detected hydrogen
      (SN~2010gx and SN~2007bi) are indicated with double pentagons.
      The contours show the distribution of 
      SDSS galactic metallicities from \citet{tremonti04}.  
      For comparison we plot the host galaxies of low-redshift long GRBs 
      (red points) showing that both they and the luminous SNe hosts appear 
      primarily in the low-metallicity tail of the galaxy 
      distribution. 
      We calculate abundances of the \citep{russell90} sample of
      \ion{H}{2} regions for the LMC and SMC calculated with the N2 
      diagnostic of \citet{pp04} and plot ranges and medians
      with cyan stars.  (We stress that the actual ranges in
      the LMC and SMC metallicities are likely lower than the ranges given by 
      applying the \citet{pp04} N2 diagnostic to these \ion{H}{2} regions; 
      there is a fairly large scatter inherent in the diagnostic.)
      Host galaxies of luminous SNe that have no measurements yet of host 
      metallicity are indicated with arrows at their absolute B magnitude.
      The right vertical axis shows the corresponding scale of oxygen 
      abundance in terms of the Solar value from \citet{delahaye06}.
      All metallicity measurements have been converted to 
      the scale of the N2 diagnostic of \citet{pp04} using the 
      conversions of \citet{kewley08}.
   }
  \end{figure}

The low metallicity of the host of SN~2010jl is another confirmation of the 
emerging trend of low-metallicity hosts of very luminous supernovae 
\citep{kozlowski10}.  In 
Figure~\ref{fig:ZBpp04n2} we show this trend for SN~2010jl (green squares) 
and other luminous SNe (blue pentagons; SN~2003ma, SN~2007bi, SN~2007va, and 
SN~2010gx).  The oxygen abundance measurement for the host of SN~2010gx 
was obtained from our own spectroscopic observations, which we present here. 
  The luminous supernovae without detected hydrogen 
(SN~2010gx and SN~2007bi) are indicated with double pentagons. 
We plot three metallicity measurements for the the host region of 
SN~2010jl with green squares:  two from our OSMOS spectra of the 
regions $\sim$1 kpc north and south of the supernova and one from 
the SDSS spectrum $\sim$1 kpc northeast of the supernova; the value from 
strong line fitting published by \citet{tremonti04}.  We cannot convert the 
\citet{pilyugin07} measurement of 8.3 to the scale of the \citet{pp04} N2
diagnostic and so to maintain a 
strictly uniform 
abundance scale we do not plot it here.  All 
four are consistent to within the errors, indicating that there 
is no 
strong metallicity variation in the area and that the strong line 
diagnostics are well-behaved and not in the poorly-calibrated 
regime of very hard radiation fields. 
For comparison, we also plot contours showing the distribution of the 
measured SDSS galaxy metallicities from \citet{tremonti04}.  
We also plot for comparison the medians and ranges of nebular abundances 
measured in the LMC and SMC with strong line diagnostics 
from \ion{H}{2} flux ratios reported by \citet{russell90} and absolute B 
magnitudes from \citet{karachentsev05}.  We stress that the actual range in 
the LMC and SMC metallicities is likely lower than the range given by applying 
the \citet{pp04} N2 diagnostic to these \ion{H}{2} regions; there is 
a fairly large scatter inherent in the diagnostic.  
We plot six low-$z$ GRB/SN hosts (red circles) to show the similarity in 
metallicity trends for luminous SNe and GRB hosts 
\citep{chornock10,levesque10a}.

For high-luminosity SNe which do not yet have direct 
host metallicity measurements, we plot arrows at the host absolute magnitudes
or at the upper limit the host absolute magnitude
(from left to right: SN~2005ap, SN~2006tf, SN~2008fz, SN~2005gj, PTF~10hgi, SN~2008es, 
PTF~10heh, PTF~09cnd, SCP06F6, PTF~10vqv, SN~2009jh, SN~1999as, PTF~09atu, 
and SN~2006gy).  By comparing the distribution of these to the 
\citet{tremonti04} contours, it is clear that the host galaxies of these 
luminous SNe are in fainter galaxies.   \citet{neill10} have also shown 
that the host galaxies of the most luminous SNe are the faintest, bluest 
galaxies.  These hosts for which the metallicity have not yet been determined 
themselves provide support for the low-metallicity trend among 
luminous SNe hosts, because the faintest galaxies tend to be the most 
metal-poor \citep[e.g.][]{tremonti04}.

\begin{figure}
  \begin{center}
  \begin{tabular}{c}
  \plottwo{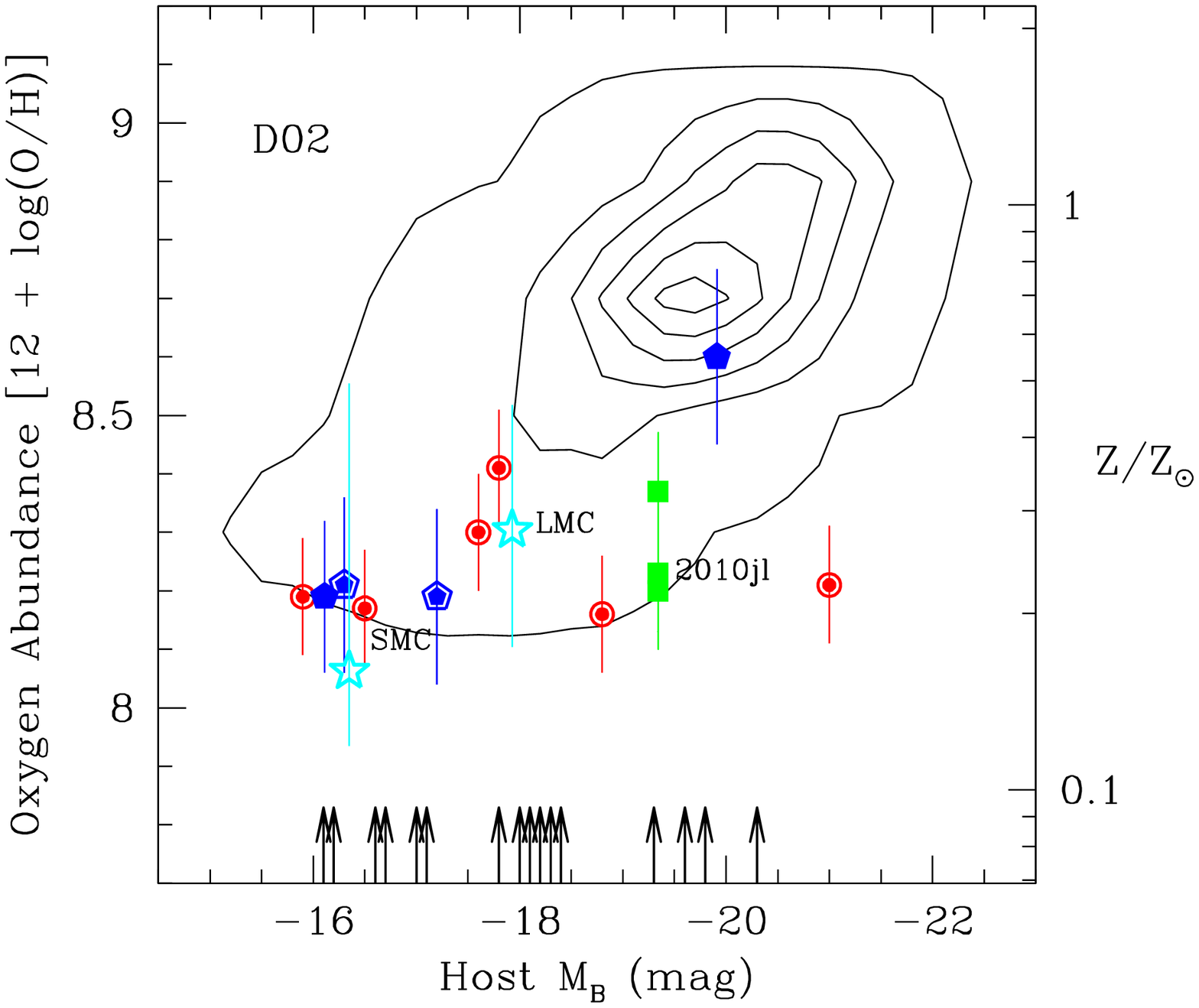}{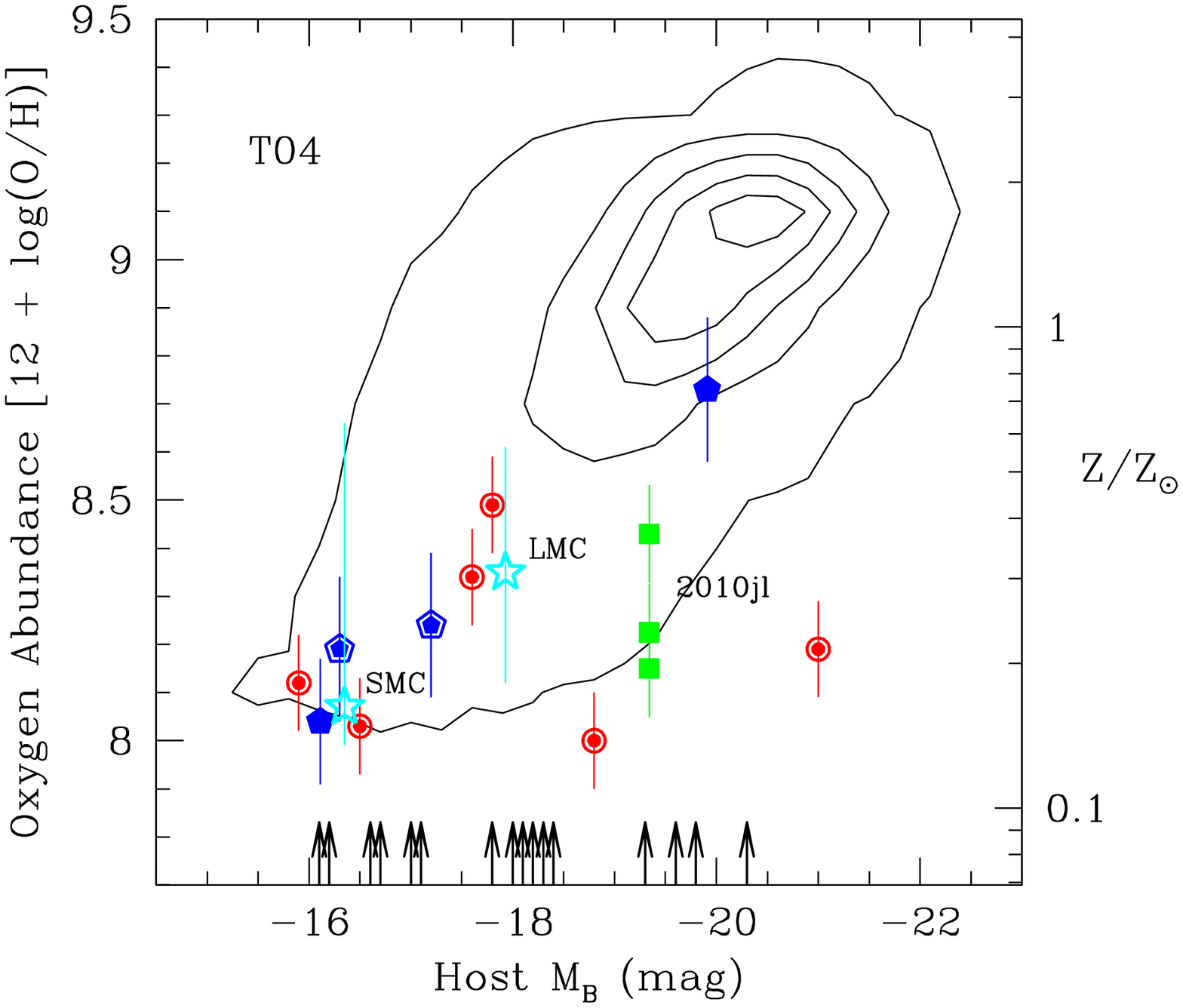}
  \end{tabular}
  \end{center}
  \caption 
   { \label{fig:ZBother}
      Same as Figure~\ref{fig:ZBpp04n2}, except (left) with all metallicities 
      on or converted to the scale of the \citet{d02} strong-line diagnostic, 
      or (right) with all metallicities on or converted to the scale of the 
      \citet{tremonti04} diagnostic.  
      We emphasize that the low-metallicity trend is not an artifact of which 
      metallicity scale one prefers.
   }
  \end{figure}

These metallicity determinations were made with a number of different 
diagnostics, and so we have converted all to a single scale using the 
empirical conversions of \citet{kewley08}.  To emphasize that the 
low-metallicity trend is not an artifact of the particular common scale we 
choose, we plot the same thing on two different scales, that of the N2 
diagnostic of \citet{pp04} in Figure~\ref{fig:ZBpp04n2} and that of 
\citet{d02} and \citet{tremonti04} in Figure~\ref{fig:ZBother}.  These were 
chosen 
to avoid a gap in the range where the \citet{kewley08} conversions from 
\citet{tremonti04} are valid.  Slightly under 2\% of the galaxies plotted 
have \citet{tremonti04} 
metallicities 
outside the 8.05--9.2 span where the \citet{kewley08} conversions are valid 
(mostly $>9.2$), and these are excluded from the contour plot; the flat 
contour at the top of each plot is due to this exclusion.  
In Figure~\ref{fig:ZBother}, we approximated the \citet{kd02} metallicity 
of GRB020903 as 8.1 instead of $8.07\pm 0.1$ in order for the \citet{kewley08}
conversion to be valid, and averaged the converted metallicities from the 
N2 scale of \citet{pp04} and \citet{d02} for the LMC, the SMC, and 
the north and south galaxy regions of the host of SN~2010jl observed by OSMOS.
On each of these three scales, the low-metallicity trend is clear.

\begin{figure}
  \begin{center}
  \begin{tabular}{c}
  \plottwo{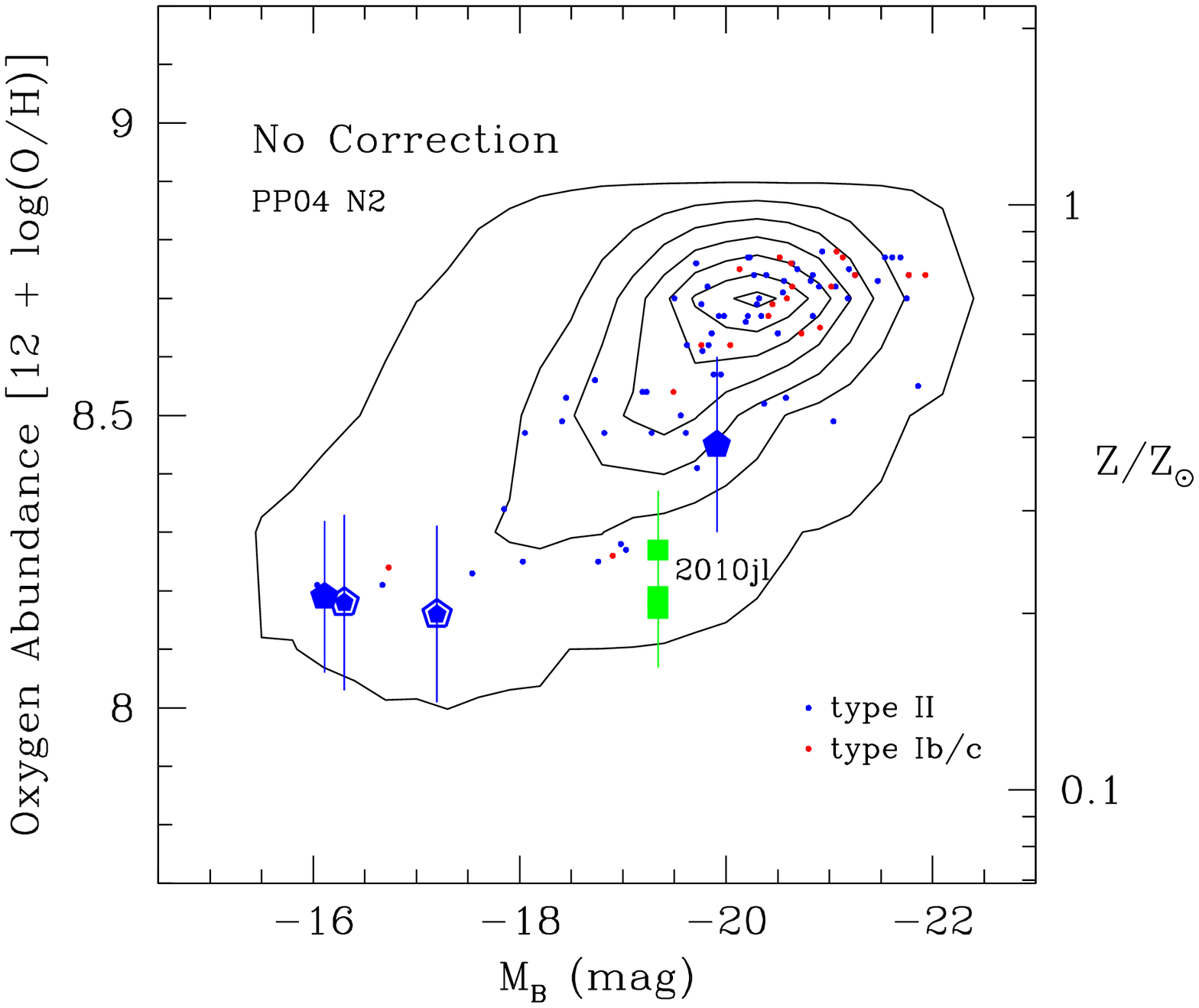}{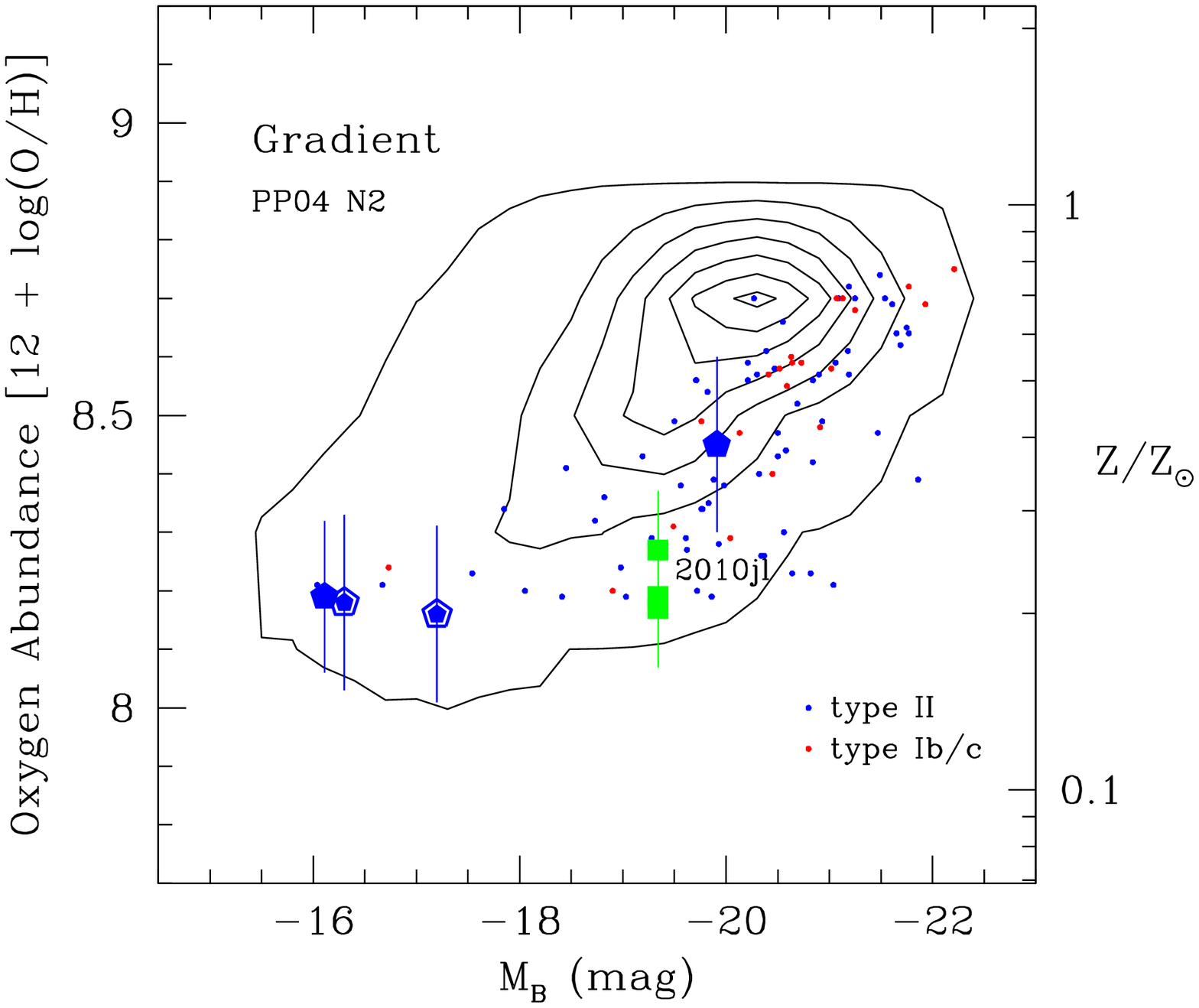}
  \end{tabular}
  \end{center}
  \caption 
   { \label{fig:GradNo}
    Type II and type Ib/c SNe that have hosts with SDSS spectra and 
    metallicities \citep{tremonti04} are plotted in the same 
    way as Fig.~\ref{fig:ZBpp04n2}.  On the left, the galactic 
    metallicities have been approximately corrected for metallicity gradients 
    to the position of the SN.  On the right, no such correction has been made.
    As in previous figures, all metallicities have been corrected to the scale 
    of the \citet{pp04} N2 diagnostic 
    using the conversions of \citet{kewley08}.  The hosts of the 
    overluminous SNe from Fig~\ref{fig:ZBpp04n2} have been plotted here again 
    for reference.
   }
  \end{figure}

\citet{prieto08} found that SN~Ib/c occur more frequently in higher-metallicity
galaxies compared to SN~II and SN~Ia.  
The highly luminous SNe we show in Figure~\ref{fig:ZBpp04n2} appear to show a 
larger bias towards low-metallicity hosts than normal SN~Ib/c.  
For comparison, in Figure~\ref{fig:GradNo} we plot host metallicities for 
type II and type Ib/c SNe that have hosts that have SDSS spectra and are 
in the redshift range $0.04 > z > 0.01$.  
We caution that these SNe are from many different surveys, each of which 
has its own selection biases.
On the left, the galactic metallicities are provided as 
observed, in the central three arcseconds of the galaxy.  
On the right, 
they have been approximately corrected to the position 
of the supernova for the expected metallicity gradient often observed in 
similar galaxies.  We use an empirical formula for the metallicity 
gradient in dex/kpc as a function of absolute $B$ magnitude of the 
host galaxy that is derived from the data published in \citet{moustakas06}.
We assume that dwarf galaxies with $M_B > -18$~mag are chemically 
homogeneous, and we assume no correction to the metallicity for supernovae 
within the 1.5$\arcsec$ radius of the SDSS fiber.

To roughly quantify the difference, we performed KS tests and found that for 
the naive (and likely incorrect) assumption of uniform selection, the 
probability of 
the luminous SNe and the type II or type Ib/c SNe progenitor regions being 
drawn from the same distribution in metallicity ranges from 
$3.9\times 10^{-4}$ to $5.6\times 10^{-5}$.  
Given the very different selection 
functions of these samples, though, this is in no way statistically rigorous, 
and should merely be taken as thought-provoking.  Some related results 
have been recently shown to hold for homogeneous and untargeted samples; 
\citet{arcavi10} show that there is a significant excess of SNe IIb in dwarf 
hosts, and that while normal SNe Ic dominate in giant galaxies, all type I 
core-collapse events in dwarf galaxies are type Ib or broad-lined type Ic.

\section{Discussion}\label{sec:discuss}

The metallicity of the hosts of SN~2010jl and SN~2010gx support and 
strengthen the emerging trend that luminous supernovae occur preferentially 
in metal-poor hosts.  
This result suggests that luminous SNe have low-metallicity 
progenitors.
It is an interesting parallel to 
the similar trend observed for the hosts of long GRBs.  
The trend is not an artifact of 
which metallicity diagnostic is chosen for the host galaxies.  

The relative iron abundance of these metal-poor hosts is expected to be 
even lower than suggested by the oxygen abundance that we use as a proxy 
for metallicity.  At low metallicity, $\alpha$-elements like oxygen are 
enhanced relative to iron, compared to the solar mixture.  
In the galactic disk and halo, at [Fe/H]~$ > -1$, 
[O/Fe] is approximately inversely proportional to [Fe/H], while below 
[Fe/H] = $-1$, the relationship flattens out at a constant (and lower) 
relative iron abundance 
\citep[e.g][]{tinsley79,mcwilliam97,johnson07}.
The iron abundance is more fundamentally important for the 
late-stage evolution of massive stars because iron provides much of the 
opacity for radiation-driven stellar winds 
\citep[e.g.][]{pauldrach86,vink05}.  
By using oxygen abundance as a metallicity tracer, we are in fact 
underestimating the magnitude of the effect of low-metallicity on 
mass-loss.  If there is a threshold in iron abundance for certain 
types of luminous SNe, the observed effect in oxygen abundance will be 
less pronounced.  The substantial preference we see for these SNe to 
occur in hosts with low nebular abundance of oxygen points to a more 
stringent constraint in iron abundance.

Nebular oxygen abundances determined by electron temperature methods 
appear to track the stellar abundances of massive young stars very 
closely, as \citet{bresolin09} demonstrate by comparing the gas-phase 
metallicity gradient from \ion{H}{2} regions to the gradient in stellar 
abundances in NGC~300, so the gas-phase oxygen abundances we discuss 
should well characterize the stellar abundances of the massive young  
progenitors of these supernovae.

A number of observational results for normal SN~Ib/c and SN~II support a 
change in their number ratios as a function of metallicity.  In an early 
study, \citet{prantzos03} used the absolute magnitudes of supernova host 
galaxies as a proxy for their average metallicities from the 
luminosity-metallicity relationship, and found that the number ratio 
of normal SN~Ib/c to SN~II increases with metallicity.  \citet{prieto08} 
took advantage of the large sample of well-observed and typed supernovae in 
the literature with star-forming host galaxies that have high S/N spectra 
obtained by the Sloan Digital Sky Survey (SDSS).  They found strong evidence 
that normal SN~Ib/c, dominated in rates by SN~Ic, occur more frequently in 
higher-metallicity galaxies compared to SN~II and SN~Ia.  Recently, 
\citet{modjaz10} found that the local metallicity of the hosts of SN~Ic is 
on average higher than for hosts of SN~Ib \citep[but see][]{anderson10}.  
All these results are consistent with 
state-of-the-art stellar evolution models of massive stars with rotation, 
either single or in binaries \citep[e.g.][]{eldridge08,georgy09,georgy10}.

Studies of local and cosmological long-duration GRBs find that their host 
galaxies have low metallicity compared to the general population of 
star-forming galaxies and other CCSNe.  These results help put constraints on 
progenitor models and the physics of production of the GRB jet, and they 
have fundamental implications for the use of GRBs as star-formation 
tracers at high redshift \citep[e.g.][]{kistler09}.  \citet{fruchter06} 
initially pointed out that the host environments of cosmological GRBs were 
different from CCSNe, with GRBs being hosted by strongly star-forming dwarf 
galaxies \citep[see also][]{lefloch03}.  \citet{stanek06} compared direct 
host metallicity measurements of five local GRBs with those of star-forming 
galaxies in SDSS, and found that long GRBs prefer low-metallicity 
environments.  This result has recently been confirmed using a larger sample 
of events \citep{levesque10b}.  \citet{modjaz08} used direct host galaxy 
metallicity estimates to study nearby broad-lined SN~Ic.  They find that the 
metallicities of hosts of SN~Ic without GRB are higher than the metallicities 
of hosts of SN~Ic associated with GRBs.

\subsection{Luminous supernovae}

One of the most surprising results of the new transient surveys that are 
monitoring many square degrees of the sky every few nights has been the 
discovery of very luminous supernovae in the nearby Universe that are 
$\sim 10$~times more luminous than normal core-collapse supernovae 
and can release $\sim 10^{51}$~ergs in UV-optical light.  

The first events of this kind were SN~2005ap \citep{quimby07} and SN~2006gy
\citep{smith07gy}, discovered by TSS. 
Their light curves, spectral properties, and 
environments are remarkably different, although they both share extreme 
energetics compared to normal CCSNe.  SN~2005ap was discovered in a 
low-luminosity galaxy at z = 0.28.  It showed a relatively fast evolution 
in its light curve, but with an extreme peak absolute magnitude of $-23$~mag 
never before seen in a supernova.  The spectra had very broad features that 
were mainly associated with CNO events.  The absence of hydrogen in the 
spectrum, the features at high velocity, and the extreme energetics led 
\citet{quimby07} to propose that SN~2005ap could be associated with long 
GRBs, which connected this event with a high-mass progenitor star.

SN~2006gy was discovered close to the center of a nearby (z = 0.02) S0 
galaxy with recent star formation.  Its spectrum and light curve 
were similar to type IIn SNe, but they were much more extreme, with a 
peak absolute magnitude of $-22$ and clear signs of interaction between the 
supernova ejecta and a massive, hydrogen-rich circumstellar shell.  This 
led \citet{smith07gy} to propose that the progenitor of SN~2006gy had been a 
very massive star (initial mass $\gtrsim 100$~M$_\odot$) similar to the 
well-studied galactic Luminous Blue Variable (LBV) $\eta$-Carinae.  
The explosion mechanism of SN~2006gy might be explained by the pulsational 
pair-instability model \citep{woosley07}.  

\citet{vanmarle10} explore possible models for luminous type IIn SNe, and 
find within the parameter space that they explore that sustained high 
luminosity requires massive circumstellar shells.  \citet{metzger10} 
consider interaction with a relic protostellar disk, finding that when the 
disk is massive and compact and the cooling time is long compared to the 
expansion timescale, more of the energy is radiated in the optical, which they 
suggest is a possible explanation of luminous type IIn events.  
\citet{murase10} explore potential consequences of such interactions with 
circumstellar shells for the production of gamma rays and neutrinos.  

If LBV-like stars are the progenitors of these extremely luminous supernovae, 
that does not constrain them to live in high-metallicity environments; 
LBVs have been identified in galaxies with $\rm 12 + log(O/H) \lesssim 8.0$ 
\citep{izotov09,herrero10}. 
The progenitor of the luminous type IIn SN~2010jl is constrained by 
\citet{smith10jl} be a massive 
LBV, a less massive object temporarily experiencing an LBV-like eruption 
in a precursor explosion, or a member of a compact, massive cluster with 
turnoff mass $\gtrsim 30$~M$_\odot$.  
The progenitors of two less luminous 
type IIn events have been identified.  One of these, SN~2005gl, occurred in 
a region with gas-phase metallicity of $\rm 12 + log(O/H) = 9.1\pm0.3$, and 
its progenitor was identified as an LBV with $M_V = -10.3$ \citep{galyam09gl}.
The other, SN~1961V, a low-metallicity, normal-luminosity type IIn 
\citep{kochanek10,smith10lbv}, had a 
progenitor with $\rm M_{ZAMS} > 80 M_\odot$, which must have been greater 
than $\sim 30$~M$_\odot$ at the time of explosion 
to retain any hydrogen \citep{kochanek10}.

Since the discovery of SN~2005ap and SN~2006gy, there have been several 
very energetic supernovae (peak $M_R < -20$) discovered by different surveys 
out to redshift 
$z \approx 1$.  We present a list in Table~\ref{table:lumSNe}; we include 
whether each event is classified as type I (without hydrogen) or type II 
(with hydrogen).  Many host absolute B magnitudes are converted and errors 
might be $\sim 0.2$~mag or larger.
The new discoveries include objects with a variety of properties.  
Some events are similar to 
SN~2005ap, including 
SCP06F6, PTF~09cnd, PTF~09atu, and SN~2009jh \citep{quimby09}, 
SN~2010gx \citep{mahabal10,quimby10gx,mahabal10err,pastorello10}, PTF~10hgi \citep{quimby10hgi}, 
and PTF~10vqv \citep{quimby10vqv}, and to SN~2006gy, including SN~2003ma 
\citep{rest09}, SN~2005gj \citep{aldering06,prieto07}, SN~2006tf \citep{smith08tf}, 
SN~2008es \citep{miller09, gezari09}, SN~2008fz \citep{drake10}, and 
PTF~10heh \citep{quimby10heh}.  Others are more similar to broad-lined 
SN~Ic in spectroscopic properties, but with extreme luminosity, such as 
SN~1999as and SN~2007bi \citep{galyam09bi}.  We also include a unique event, 
SN~2007va, that was completely enshrouded in its own circumstellar dust 
and only observable at mid-infrared wavelengths; its infrared luminosity 
indicated high reprocessed optical luminosity \citep{kozlowski10}.  
The most recent example (and nearest, at $z=0.011$) is the type IIn 
SN~2010jl, the main subject of this paper.  
The variety of observed properties in these luminous CCSNe likely represents 
different explosion mechanisms (e.g.~normal collapse of the core vs. 
pair instability vs. pulsational pair-instability), different mechanisms 
that power the energetics of the light curves (e.g.~ circumstellar 
interaction vs. radioactive decay vs. magnetar spindown), and different 
progenitor mass-loss histories and masses (e.g.~LBV vs. Wolf-Rayet stars vs.
red supergiants).

\begin{deluxetable}{lcccccc}
\tablecolumns{7}
\tabletypesize{\scriptsize}
\tablewidth{0pt}
\tablecaption{List of luminous CCSNe collected from the literature\label{table:lumSNe}}
\tablehead{
   \colhead{SN Name}                &
   \colhead{RA}                     &
   \colhead{Dec}                    &
   \colhead{SN Spectral}            &
   \colhead{Redshift}               &
   \colhead{$M_B$(host)}            &
   \colhead{Host Metallicity?}      \\
   \colhead{}                       &
   \colhead{(J2000.0)}              &
   \colhead{(J2000.0)}              &
   \colhead{Type}                   &
   \colhead{}                       &
   \colhead{(mag)}                  &
   \colhead{(12+log[O/H], PP04N2)}          
}
\startdata
SN 1995av & 02:01:36.7 & 03:38:55 & IIn & 0.3    & \nodata &  No \\ 
SN 1997cy & 04:32:54.8 & --61:42:57 & IIn & 0.063 & $-17.8$    &  No \\ 
SN 1999as & 09:16:30.8 & +13:39:02 & Ic & 0.13 & $-18.4$ & No \\ 
SN 1999bd & 09:30:29.2 & +16:26:08 & IIn & 0.151 &  $-18.4$    & No \\ 
SN 2000ei & 04:17:07.2 & +05:45:53 & IIn & 0.600 & $>-19.3$      & No \\ 
SN 2003ma & 05:31:01.9 & --70:04:16 & IIn  & 0.29 & $-19.9$ & $8.45$ \\ 
SN 2005ap & 13:01:14.8 & +27:43:31 & Ic?  & 0.28 & $-16.1$ & No \\ 
SN 2005gj & 03:01:12.0 & +00:33:14 & Ia/IIn & 0.06 & -16.7 & No \\
SCP06F6   & 14:32:27.4 & +33:32:25 & Ic?  & 1.19 & $> -18.1$ & No \\ 
SN 2006gy & 03:17:27.0 & +41:24:20 & IIn  & 0.02 & $-20.3$ & No \\
SN 2006tf & 11:56:49.1 & +54:27:26 & IIn  & 0.07 & $-16.2$ & No \\
SN 2007bi & 13:19:20.2 & +08:55:44 & Ic   & 0.13 & $-16.3$ & $8.18$ \\
SN 2007va & 14:26:23.2 & +35:35:29 & IIn? & 0.19 & $-16.2$ & $8.19$ \\
SN 2008am & 12:28:36.2 & +15:34:49 & IIn  & 0.234 & $-19.6$  &  No \\
SN 2008es & 12:46:15.8 & +11:25:56 & IIL & 0.21 & $> -17.1$ & No \\
SN 2008fz & 23:16:16.6 & +11:42:48 & IIn & 0.13 & $> -16.6$ & No \\
PTF 09atu & 16:30:24.6 & +23:38:25 & Ic? & 0.50 &   $> -19.8$ & No \\
PTF 09cnd & 16:12:08.9 & +51:29:16 & Ic? & 0.26 &   $> -18.1$ & No \\
SN 2009jh & 14:49:10.1 & +29:25:11 & Ic? & 0.35 &   $> -18.3$ & No \\
SN 2010gx & 11:25:46.7 & --08:49:41 & Ic? & 0.23 &  $-17.2$ & $8.16$ \\
PTF 10heh & 12:48:52.0 & +13:26:24.5 & IIn & 0.34 & $-18.0$ & No \\
PTF 10hgi & 16:37:47.0 & +06:12:32.3 & Ic? & 0.10 & $> -17.0$ & No \\
PTF 10vqv & 03:03:06.8 & --01:32:34.9 & Ic? & 0.45 & $-18.2$ & No \\
SN 2010jl & 09:42:53.3 & +09:29:41.8 & IIn & 0.01 & $-19.3$ & $8.19$ \\
\enddata
\end{deluxetable}

\subsection{Metallicity diagnostics}\label{sec:zdiag}

The forbidden and recombination lines in the optical spectra of star-forming 
galaxies provide the observables needed to estimate chemical abundances, 
star-formation rates, and other physical properties of the \ion{H}{2} 
regions, like 
electron temperature and density.  These emission lines are produced by 
collisional excitation and recombination processes in gaseous nebulae where 
massive ($\gtrsim 20$~M$_\odot$) stars form, thanks to the interactions of 
the ionizing UV photons provided by the stars and the surrounding gas 
\citep[e.g.][]{osterbrock06}.

There is a vast literature on a number of techniques that have been developed 
over the last three decades to estimate the oxygen abundances and physical 
conditions of \ion{H}{2} regions in star-forming galaxies 
\citep[e.g.][and references therein]{kewley08}.  The methods can be divided 
into four different categories:  direct (uses an estimate of the electron 
temperature measured from the faint [\ion{O}{3}]$\lambda 4363$\AA{} auroral line),
empirical \citep[e.g.][]{pp04}, theoretical \citep[e.g.][]{kk04}, and a 
combination of empirical and theoretical \citep[e.g.][]{d02}.  All these 
methods are based on high S/N measurements of the line ratios of different 
combinations of emission lines from ions present in the optical region of 
the spectrum ($\simeq 3700-6700$\AA): 
[\ion{O}{2}], [\ion{O}{3}], [\ion{N}{2}], [\ion{S}{2}], 
H$\alpha$, and H$\beta$.

Each method has relative advantages and shortcomings that have been 
investigated in a number of studies in the literature 
\citep[e.g.][]{yin07,moustakas10}, and there are known discrepancies of 
as much as $\sim 0.6$~dex between different estimates of the oxygen 
abundances.  For example, the estimated oxygen 
abundance of the host galaxy of the luminous SN~2007va \citep{kozlowski10} 
obtained from the direct electron temperature method is 0.3 dex lower than 
the value obtained from the theoretical calibration of 
$R_{23} =$~([\ion{O}{2}]$\lambda 3727 +$[\ion{O}{3}]$\lambda\lambda 
4959, 5007$)/H$_\beta$ \citep{kk04}.  \citet{bresolin09} demonstrate the 
offsets between the various techniques for their determination of the 
metallicity gradient in NGC~300.  The abundances estimated with one 
diagnostic, the N2 diagnostic of \citet{pp04}, essentially track the stellar 
and ($T_e$) nebular abundances they measure, albeit with rather more 
scatter. 
Because the effective zero points and scales vary somewhat, regardless of 
which method is used, it is important to use (or convert to) just one.

With the S/N and wavelength coverage of our observations, we content ourselves 
in this paper with the empirical O3N2 and N2 diagnostics of \citet{pp04} and 
the theoretical and empirical N2 diagnostic of \citet{d02}.  We use the 
empirical conversions of \citet{kewley08} to convert metallicities  
determined on other scales to these for the purpose of direct comparison.

\section{Conclusions}

Recent observations of the environments 
of supernovae and GRBs show that metallicity is a key parameter in the lives 
and deaths of massive stars. 
Our finding that the hosts of luminous SNe lie in the low-metallicity 
extremes of the distribution of star-forming galaxies supports this  
emerging picture.  
Verifying and constraining this emerging 
relationship with metallicity may be an important probe of the mechanisms of 
the most luminous supernovae, and this trend may be an important factor in 
various aspects of the evolution of the metal-poor early universe.


\acknowledgments

We would like to thank Andrew Drake, Avishay Gal-Yam, Jennifer Johnson, 
Rubab Khan, Chris Kochanek, Paul Martini, Brian Metzger, Nidia Morrell, 
Rupak Roy, Josh Simon, Firoza Sutaria, and the anonymous referee for 
discussion, comments, and assistance.  
JLP acknowledges support from NASA through Hubble Fellowship grant 
HF-51261.01-A awarded by the STScI, which is operated by AURA, Inc. for 
NASA, under contract NAS 5-26555.  RS is supported by the David 
G. Price Fellowship in Astronomical Instrumentation.  GP is supported 
by the Polish MNiSW grant N203 007 31/1328.  OSMOS has been 
generously funded by the National Science Foundation (AST-0705170) and 
the Center for Cosmology and AstroParticle Physics at The Ohio State 
University.
Funding for the SDSS and SDSS-II has been provided 
by the Alfred P. Sloan Foundation, the Participating Institutions, the 
National Science Foundation, the U.S. Department of Energy, the National 
Aeronautics and Space Administration, the Japanese Monbukagakusho, the 
Max Planck Society, and the Higher Education Funding Council for England.
The SDSS Web Site is http://www.sdss.org/.  
The SDSS is managed by the Astrophysical Research Consortium for the 
Participating Institutions.  The Participating Institutions are the 
American Museum of Natural History, Astrophysical Institute Potsdam, 
University of Basel, University of Cambridge, Case Western Reserve University, 
University of Chicago, Drexel University, Fermilab, the Institute for 
Advanced Study, the Japan Participation Group, Johns Hopkins University, 
the Joint Institute for Nuclear Astrophysics, the Kavli Institute for 
Particle Astrophysics and Cosmology, the Korean Scientist Group, the 
Chinese Academy of Sciences (LAMOST), Los Alamos National Laboratory, the 
Max-Planck-Institute for Astronomy (MPIA), the Max-Planck-Institute for 
Astrophysics (MPA), New Mexico State University, Ohio State University, 
University of Pittsburgh, University of Portsmouth, Princeton University, 
the United States Naval Observatory, and the University of Washington.
This research has made use of the NASA/IPAC Extragalactic Database (NED) 
which is operated by the Jet Propulsion Laboratory, California Institute of 
Technology, under contract with the National Aeronautics and Space 
Administration.

Version tracking:  This draft \TeX{}ed on \today.

{\it Facilities:} \facility{Hiltner (OSMOS), \facility{Sloan}, \facility{Du Pont (WFCCD)}, \facility{Swope}}.

\bibliography{SNe} 
\bibliographystyle{apj} 

\end{document}